\documentclass[aps,prb,twocolumn,floatfix,preprintnumbers,superscriptaddress,amsmath,amssymb,longbibliography]{revtex4-2}
\usepackage{xcolor,bm,lineno,multirow,float,times}
\usepackage{physics}
\usepackage{rotating}
\usepackage{verbatim}
\usepackage{graphicx,tabularx}
\usepackage[multiple]{footmisc}
\usepackage[sort&compress]{natbib}
\usepackage[T1]{fontenc}
\usepackage{soul}

\usepackage[%
  colorlinks=true,
  urlcolor=blue,
  linkcolor=blue,
  citecolor=blue
]{hyperref}

\newcommand{\blue}[1]{{\color{black} #1}}

\def\ie{{i.e. }}

\begin{document}

\title{Electronic and magnetic properties of intermetallic  Kagome magnets $R$V$_6$Sn$_6$ ($R$ = Tb - Tm)}

\author{X. X. Zhang}
\affiliation{Beijing National Laboratory for Condensed Matter Physics and Institute of Physics, Chinese Academy of Sciences, Beijing 100190, China}
\affiliation{School of Physical Sciences, University of Chinese Academy of Sciences, Beijing 100190, China}

\author{Z. Y. Liu}
\affiliation{Beijing National Laboratory for Condensed Matter Physics and Institute of Physics, Chinese Academy of Sciences, Beijing 100190, China}
\affiliation{School of Physical Sciences, University of Chinese Academy of Sciences, Beijing 100190, China}

\author{Q. Cui}
\affiliation{Beijing National Laboratory for Condensed Matter Physics and Institute of Physics, Chinese Academy of Sciences, Beijing 100190, China}
\affiliation{School of Physical Sciences, University of Chinese Academy of Sciences, Beijing 100190, China}

\author{Q. Guo}
\affiliation{Beijing National Laboratory for Condensed Matter Physics and Institute of Physics, Chinese Academy of Sciences, Beijing 100190, China}
\affiliation{School of Physical Sciences, University of Chinese Academy of Sciences, Beijing 100190, China}

\author{N. N. Wang}
\affiliation{Beijing National Laboratory for Condensed Matter Physics and Institute of Physics, Chinese Academy of Sciences, Beijing 100190, China}
\affiliation{School of Physical Sciences, University of Chinese Academy of Sciences, Beijing 100190, China}

\author{L. F. Shi}
\affiliation{Beijing National Laboratory for Condensed Matter Physics and Institute of Physics, Chinese Academy of Sciences, Beijing 100190, China}
\affiliation{School of Physical Sciences, University of Chinese Academy of Sciences, Beijing 100190, China}

\author{H. Zhang}
\affiliation{Beijing National Laboratory for Condensed Matter Physics and Institute of Physics, Chinese Academy of Sciences, Beijing 100190, China}

\author{W. H. Wang}
\affiliation{Beijing National Laboratory for Condensed Matter Physics and Institute of Physics, Chinese Academy of Sciences, Beijing 100190, China}
\affiliation{Songshan Lake Materials Laboratory, Dongguan, Guangdong 523808, China}

\author{X. L. Dong}
\affiliation{Beijing National Laboratory for Condensed Matter Physics and Institute of Physics, Chinese Academy of Sciences, Beijing 100190, China}
\affiliation{School of Physical Sciences, University of Chinese Academy of Sciences, Beijing 100190, China}
\affiliation{Songshan Lake Materials Laboratory, Dongguan, Guangdong 523808, China}

\author{J. P. Sun}
\affiliation{Beijing National Laboratory for Condensed Matter Physics and Institute of Physics, Chinese Academy of Sciences, Beijing 100190, China}
\affiliation{School of Physical Sciences, University of Chinese Academy of Sciences, Beijing 100190, China}

\author{Z. L. Dun}
\email[]{dun@iphy.ac.cn}
\affiliation{Beijing National Laboratory for Condensed Matter Physics and Institute of Physics, Chinese Academy of Sciences, Beijing 100190, China}
\affiliation{School of Physical Sciences, University of Chinese Academy of Sciences, Beijing 100190, China}

\author{J. G. Cheng}
\email[]{jgcheng@iphy.ac.cn}
\affiliation{Beijing National Laboratory for Condensed Matter Physics and Institute of Physics, Chinese Academy of Sciences, Beijing 100190, China}
\affiliation{School of Physical Sciences, University of Chinese Academy of Sciences, Beijing 100190, China}

\date{\today}

\begin{abstract}
We present a systematic study of the structure, electronic, and magnetic properties of a new branch of intermetalllic compounds, $R$V$_6$Sn$_6$ ($R$ = Tb - Tm) by using X-ray diffraction, magnetic susceptibility, magnetization,  electrical transport, and heat-capacity measurements. 
These compounds feature a combination of a non-magnetic vanadium kagome sublattice and a magnetic rare-earth triangular sublattice that supports various spin anisotropies based on different $R$ ions. 
We find magnetic orders for the  $R$ = Tb, Dy, and Ho compounds at 4.4, 3, 2.5 K, respectively, while no ordering is detected down to 0.4 K for the  $R$ = Er and Tm compounds with easy-plane anisotropies.
Electronically, we found no superconductivity or charge ordering transition down to 0.4 K for any member of this family, while all compounds exhibit multi-band transport properties that originate from the band topology of the vanadium kagome sublattice. 
\end{abstract} 

\maketitle

{\centerline{\textbf{INTRODUCTION}}}
~\\


Kagome metals have been at the forefront of condensed matter physics due to the quantum-level interplay between geometry, topology and correlation \cite{reviewofback,zhou2017QSLreview, keimer2017physics, he2018topological, yin2018kagomemagnet, chang2013,xu2015CsLiMnF, pereiro2014topokagome, li2019FeSn, lzy2020CoSnS, cuiqi2020DyPtO, meng2010QSL}. On one hand, local moments on a lattice formed by corner-shared triangles induce strong frustration which serves as an important ingredient in realizing quantum spin liquids \cite{zhou2017QSLreview, meng2010QSL, norman2016searchforQSL, lawler2008gapless}.
On the other hand, the electronic band structure of Kagome lattice usually gives rise to flat bands, inflection points, and Dirac cones that promote non-trivial topology \cite{ghimire2020kagometopology, li2018flatband, ye2018naturedirac, zhang2017topological, li2021dirac, yin2020STM, pereiro2014topokagome,li2019FeSn,he2018topological}
The combination of these effects usually gives rise to exotic states with possible capabilities of magnetic field and high-pressure engineering.

As an example, a recently discovered family of Kagome metals,  $A$V$_3$Sb$_5$ ($A$ = K, Rb, Cs), has attracted tremendous research interest as a novel platform to study the interplay between nontrivial band topology, superconductivity , and charge density-wave (CDW) order\cite{PRMCsVSb,PRMKVSb,RbVSb,PhysRevLettsurface}. The most prominent feature of this structure is the presence of a kagome net of vanadium atoms that are coordinated by Sb atoms, giving rise to $Z_2$ topological states with Dirac nodal points near the Fermi level \cite{PRMCsVSb,PhysRevLettsurface, RbVSb}. 
Meanwhile, superconductivity was discovered at ambient pressure below 0.93, 0.92, and 2.5 K for $A$ = K, Rb, and Cs compounds, respectively\cite{PRMKVSb,RbVSb,PhysRevLettsurface}, which was found to compete with an unusual charge order at high temperature \cite{chenky2021CsVSb,yu2021unusual}. This example shows the urgent need to explore the unusual superconductivity of other vanadium-based Kagome intermetallics. 

\begin{figure*}[tbp!]
	\linespread{1}
	\par
	\begin{center}
		\includegraphics[width= 7 in]{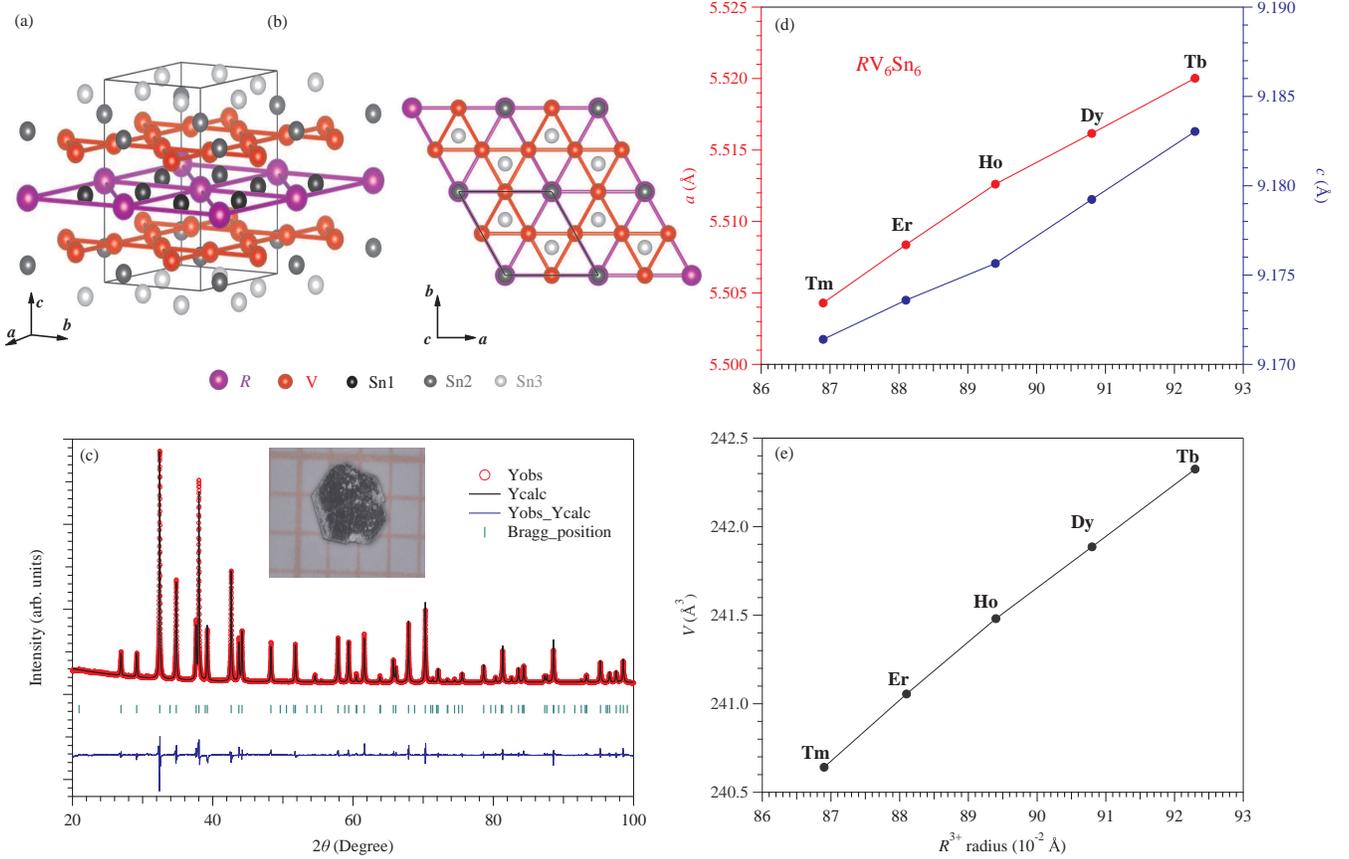}
	\end{center}
	\par
	\caption{\label{Fig:structure} (a) Crystal structure of $R$V$_6$Sn$_6$ showing alternating Kagome and triangular layers. (b) Top view of crystal structure from the $c$-axis. (c) Observed (red circle), calculated (black line), and difference (blue line) profiles of the powder XRD patterns of DyV$_6$Sn$_6$ from Rietveld refinements. Inset of (c) shows the photo of  a mm-size DyV$_6$Sn$_6$ single crystal.  Ionic radii dependence of (d) cell parameters $a$, $c$ , and (e)  the unit cell volume.}
\end{figure*}

Another large family of Kagome metal is  $RM_6X_6$ which crystallize in the MgFe$_6$Ge$_6$ structural prototype. The $R$-site hosts a variety of  rare-earth \blue{ions (Y, Gd-Lu)},  $M$ is 3d transition metal elements (e.g. Co, Cr, Mn, V, Ni...), and the X-site is generally restricted to the group IV elements (Si, Ge, Sn). In these compounds, $M$ atoms form a Kagome lattice and $R$-atoms form a triangular lattice; the two sublattices are stacked along the $c$-axis \blue{in the ABA sequence } to form a hexagonal structure [Fig.~\ref{Fig:structure}(a)(b)].  Among them, $R$Mn$_6$Sn$_6$  has recently received the most attention due to the coexistence of topological band structure and magnetic order \cite{ma2021PRLMn-based}. With the Mn-sublattice ordering at room temperature, the spin-orbit coupling and the ferromagnetic moment outside the $ab$-plane open an energy gap near the Fermi surface, giving rise to Chern gapped Dirac fermion properties \cite{ma2021PRLMn-based,yin2020NATURETb}. An important feature of the system is that both the Mn and $R$ sublattices possess localized magnetic moments which are strongly coupled with each other. Recent study has demonstrated a close relationship between rare-earth magnetism and topological electron structure, indicating that the rare-earth elements can effectively engineer the Chern quantum phase in these materials \cite{ma2021PRLMn-based,gao2021AHE}. This is partially reflected in the distinct Mn-$R$ spin orientations with different $R$ sublattice. The magnetic anisotropy varies from easy-plane for $R$ = Gd, easy-axis for $R$ = Tb, to a conical magnetic structure for $R$ = Dy and Ho \cite{venturini1991magnetic,malaman1999magnetic,clatterbuck1999magnetic}. When $R$ = Er, Tm, the Mn and rare-earth sublattices order independently in an AFM manner because the strength of the magnetic coupling is weak  \cite{malaman1999magnetic,clatterbuck1999magnetic}. \blue{In this sense, further understanding of the electronic and magnetic properties in $R$Mn$_6$Sn$_6$ requires an in-depth understanding of the $R$-sublattice magnetism. }

\begin{figure}[tbp!]
	\linespread{1}
	\par
	\begin{center}
		\includegraphics[width= \columnwidth ]{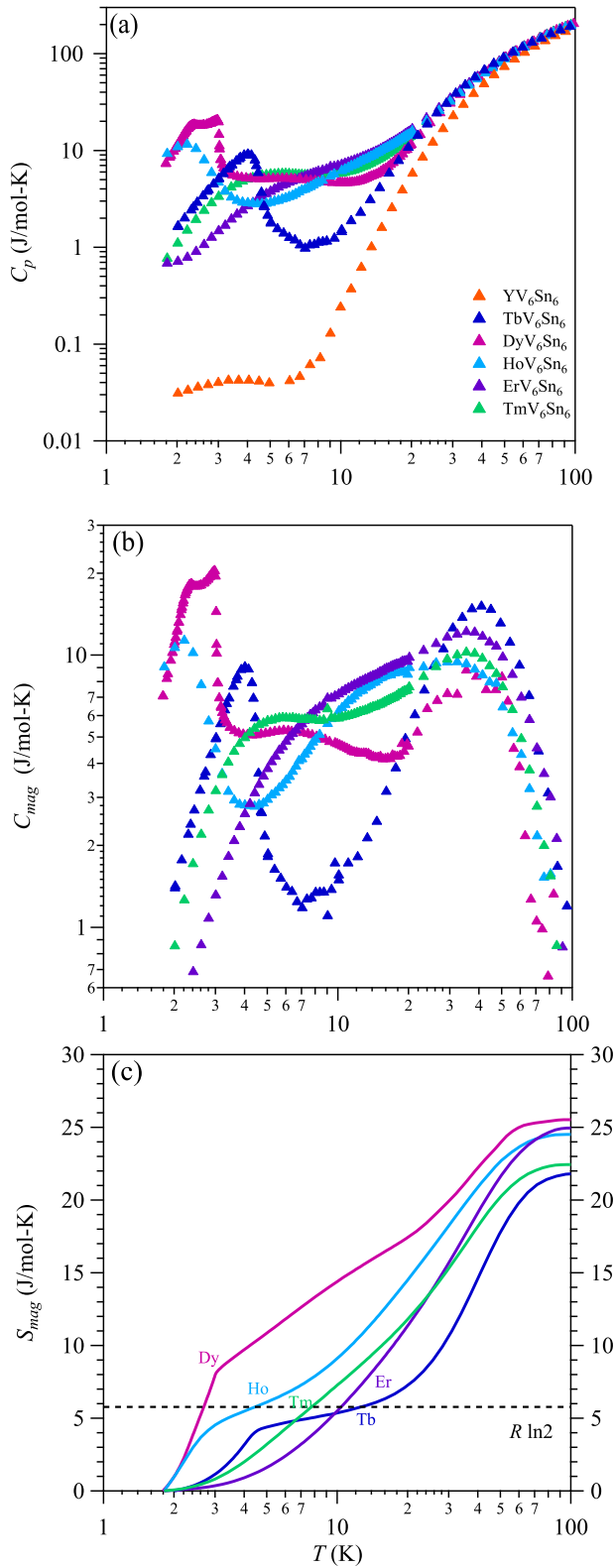}
	\end{center}
	\par
	\caption{\label{Fig:HC} \blue{(a) Measured $C_p$ for $R$V$_6$Sn$_6$($R$ = Y, Tb - Tm). (b)  Magnetic heat capacity, $C_\textrm{mag}$, obtained by subtracting the lattice contribution from $C_p$. (c) Magnetic entropy obtained from integrating $C_\textrm{mag}/T$ in (b). }  }
\end{figure}

\begin{figure*}[tbp!]
	\linespread{1}
	\par
	\begin{center}
		\includegraphics[width= 7 in]{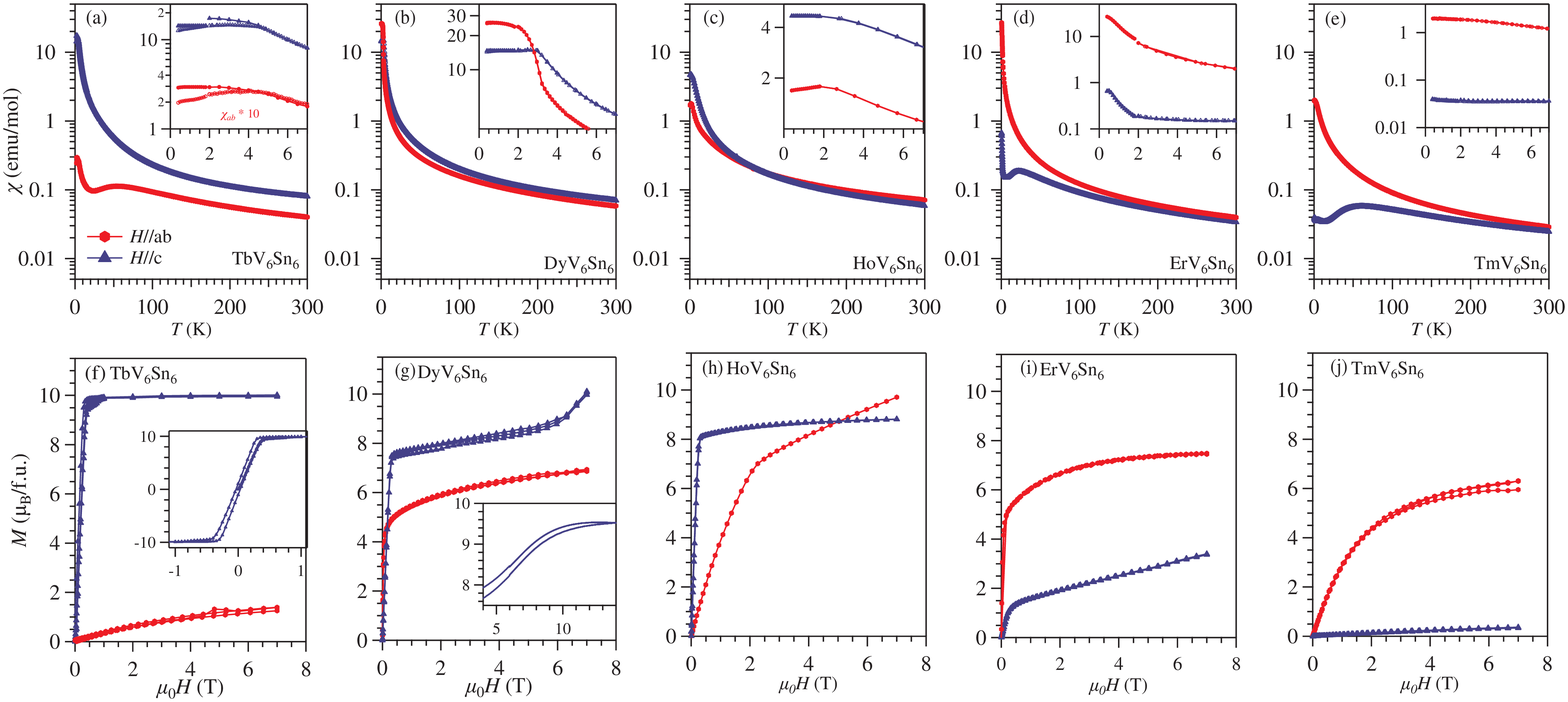}
	\end{center}
	\par
	\caption{\label{Fig:Magpro} \blue{(a-e) Magnetic susceptibility for $R$V$_6$Sn$_6$ ($R$ = Tb-Tm) in a log-linear scale.} An external magnetic field of 0.1 T was applied applied both parallel (blue triangle) and perpendicular (red circle) to the $c$-axis. Insets: zoom-in data at low temperature from 0.4 to 7 K.   (f - j) Field-dependence of isothermal magnetization measured at 0.4 K, with field applied parallel (blue triangle) and perpendicular (red circle) to the $c$-axis. Inset:  $M(H)$ curves in extended regions when hysteresis loop is present. }
\end{figure*}

In this work, we explore a new family of V-based Kagome metals $R$V$_6$Sn$_6$, and systematically study the structural, electronic, and magnetic properties for  $R$ = Gd-Tm compounds. \blue{These compounds are isostructural to its $R$Mn$_6$Sn$_6$ cousin and possess a similar non-magnetic Vanadium Kagome sublattice as $A$V$_3$Sb$_5$.  Four members of this family have been investigated very recently. Specifically,} studies have identified 2D  Kagome surface states  in HoV$_6$Sn$_6$ and GdV$_6$Sn$_6$ \cite{peng2021GdHo,pokharel2021YGd}, quantum oscillation in YV$_6$Sn$_6$\cite{pokharel2021YGd}, and CDW transition at 92 K in ScV$_6$Sn$_6$\cite{Sc}. Compared to $R$Mn$_6$Sn$_6$, \blue{the absence of}  Mn-Mn and Mn-$R$ couplings enables us to study the intrinsic rare-earth magnetism on a frustrated triangular lattice. By combining the experimental probes of x-ray diffraction (XRD), DC magnetic susceptibility [$\chi(T)$] and isothermal magnetization [$M$($H$)], heat capacity [$C_p$($T$)], \blue{as well as transport measurements including resistivity[$\rho(T)$], magneto-resistance [$MR(H)$], Hall resistivity[$\rho_{xy}$($H$)] , we show that (i) there is no superconductivity or charge ordering transition down to 0.4 K for any member of this family; (ii) all compounds exhibit multi-band transport properties that originate from the band topology of the vanadium kagome sublattice; (iii) with either strong ($R$ = Tb) or weak ($R$ = Dy, Ho) easy-axis anisotropies, the system orders at 4.4, 3, 2.5 K, respectively; (iv) with easy-plane anisotropy,  no magnetic ordering is detected down to 0.4 K for the  $R$ = Er and Tm compounds, implying the presence of magnetic frustration.}

~\\
~\\
{\centerline{\textbf{EXPERIMENTAL DETAILS}}}
~\\
~\\
Single crystals of $R$V$_6$Sn$_6$ ($R$ = Tb, Dy, Ho, Er, Tm) were synthesized via a self-flux method. Powder forms of rare-earth elements, abraded from metal blocks ( 99.99$\%$), along with V (powder, 99.9$\%$), Sn (shot, 99.999$\%$) were loaded inside an alumina crucible with the molar ratio of 1:6:50 and \blue{then sealed in evacuated quartz tubes under 10$^{-4}$ torr pressure}. The tubes were heated to 1125 $^\circ$C \blue{and dwell for 24 hours before cooling down slowly} at a rate of 2 $^\circ$C/h. The single crystals were separated from the flux via centrifuging at 825 $^\circ$C.  Crystals grown via this method were generally a few millimeters in length and 1 mm in thickness [Fig.~\ref{Fig:structure}(c) inset]. The separated single crystals were subsequently cleaned with dilute HCl to remove the flux contamination. 

Single-crystal XRD measurement on TmV$_6$Sn$_6$  were carried out on a Bruker D8 Venture single-crystal diffractometer. \blue{XYZ-centroids of 3012 reflections were collected and} integrated using the Bruker SAINT software package . Powder XRD \blue{measurements on carefully grounded single-crystal samples} were performed using a HUBER diffractometer at room temperature.  Rietveld refinements were performed with the FULLPROF software package. 

The  magnetic properties, \blue{including DC susceptibility, and isothermal magnetization}, were measured using a commercial magnetic properties measurement system (MPMS-III, Quantum Design) in the temperature range between 2 K to 300 K under different external magnetic fields. Measurements from 0.4 to 1.8 K were performed using the same MPMS with the He3 option installed. \blue{Data measured using an empty holder was used as background to correct the  diamagnetic signals in different sample environments at high and low temperatures.  With the corrected data, a modified Curie-Weiss(C-W) fit was perform from 100 to 300 K, \ie $\chi (T) = \chi_0 + C/(T-\theta_{CW})$, where$\chi_0$ denotes the temperature-independent term arising from the Pauli and van vleck  paramagnetism as well as diamagnetic signals of nucleus, and $\theta_{CW}$ is the C-W temperature.   } 

The resistivity  and specific-heat data in the temperature range from 1.8 K to 300 K were collected with a physical properties measurement system(PPMS , Quantum Design). \blue{A specific-heat measurement on a nonmagnetic YV$_6$Sb$_6$ single crystal was also measured and scaled to match that of $R$V$_6$Sb$_6$ above 100 K, which is used as an estimate of lattice contribution $C_{lat}$. The magnetic heat capacity, $C_\textrm{mag}(T)$ was then obtained by subtracting the $C_{lat}$ from the measured $C_p(T)$.} 

\begin{table}[tbp]
\centering
\caption{ \blue{Fractional atomic coordinates and equivalent isotropic displacement parameters (\AA$^2$) for TmV$_6$Sn$_6$ from the refinement of single crystal XRD data at $T$ = 273 K.} U$_{eq}$ is defined as 1/3 of of the trace of the orthogonalised U$_{ij}$ tensor.  }
\label{table:SinglecrystalXRD}
\setlength{\arrayrulewidth}{0.5mm}
\renewcommand{\arraystretch}{1.5}
\begin{tabular}{ cccccc  } 
\hline
Atom & Wyckoff pos. & $x$& $y$ & $z$ & U$_{eq}$  \\ \hline
Tm & 1$a$ &$0$& $0$& $0.5$& $0.053(3)$ \\ 
V  & 6$i$  &$0.5$ &$0.5$& $0.748016(6)$ &$0.0032(3)$  \\ 
Sn1 & 2$e$  & $0$ &$0$ &$0.83167(5)$& $0.050(3)$ \\ 
Sn2 & 2$d$  & $0.66667$& $0.33333$ &$0.5$& $0.0029(3)$ \\ 
Sn3 & 2$c$  &$0.33333$ &$0.66667$ &$0$ &$0.0037(3)$ \\ \hline
\multicolumn{6}{c}{Reflections collected: 3012} \\
\multicolumn{6}{c}{R$_1$: 2.54 \%, wR$_2$: 5.57 \% } \\
\hline
\end{tabular}
\end{table}
~\\
~\\
{\centerline{\textbf{CHARACTERIZATIONS}}}
\section{STRUCTURE} 
\blue{Powder XRD refinements confirm that all $R$V$_6$Sn$_6$ compounds crystallize in the hexagonal HfFe$_6$Ge$_6$-type structure with  space group $P6/m m m$. }
The crystallographic parameters of TmV$_6$Sn$_6$ from single-crystal XRD are listed in Table \ref{table:SinglecrystalXRD}. 
As illustrated in Fig. \ref{Fig:structure} (a, b), the ideal Kagome layers of V-ions coordinated by Sn are separated by two $R$-triangle layers. With triangle layers of rare-earth elements and Kagome layers of V elements stacked along the $c$-axis, this series of compounds exhibits the same structure as $R$Mn$_6$Sn$_6$.  
\blue{As expected}, the cell parameters $a$, $c$, as well as volume of unit cell \blue{increase monotonically}  with the increase of $R^{3+}$ ionic radii [Fig. \ref{Fig:structure} (d, e)]. 
\blue{These parameters are summarized in Table \ref{table:all}.}

\begin{table*}[htbp]
\centering
\caption{Fitted parameters for $R$V$_6$Sn$_6$ ($R$ = Tb - Tm), including (i) lattice parameters: $a$, $c$; (ii) magnetic property parameters: in-plane and out-of-plane C-W temperature, $\theta_{CW}^{ab}$, and $\theta_{CW}^{c}$, in-plane and out-of-plane effective moment, $\mu_\mathrm{eff}^{ab}$ and $\mu_\mathrm{eff}^{c}$.  }
\label{table:all}
\setlength{\arrayrulewidth}{0.3mm}
\setlength{\tabcolsep}{8pt}
\renewcommand{\arraystretch}{1.5}
\begin{tabular}{ cccccc  } 
\hline
 & TbV$_6$Sn$_6$&DyV$_6$Sn$_6$&HoV$_6$Sn$_6$&ErV$_6$Sn$_6$&TmV$_6$Sn$_6$\\ \hline
 \multicolumn{6}{c}{Lattice parameters} \\ 
 $a($\AA$)$ &5.52  & 5.5162 &5.5126  & 5.5084 & 5.5043 \\
 $c($\AA$)$ & 9.183 &9.1792  & 9.1756 & 9.1736 &9.1714  \\
 \hline
\multicolumn{6}{c}{Magnetic property parameters} \\
Spin anisotropy & easy-axis (Ising) & weak easy-axis & weak easy-axis & easy-plane & easy-plane (XY)  \\
$T_{N,C}$(K) & 4.4 &3 & 2.5 & -- & -- \\
Low Temp. $\theta_{W}^{ab}$(K)&  --& 1.29 & -10.73 & 0.07 & -3.18  \\
Low Temp. $\theta_W^c$(K).    & 2.06 & 1.65 & 0.66 & -- & --  \\ 
High Temp. $\mu_\textrm{eff}^{ab}$($\mu_B$)& 9.84&10.43&10.33&9.59&7.95 \\ 
High Temp. $\mu_\textrm{eff}^c$($\mu_B$)& 9.47&10.88&10.29&9.75&7.49 \\
$\mu_\textrm{theory}$($\mu_B$)& 9.72&10.63&10.61&9.59&7.56\\ 
\hline

\end{tabular}
\end{table*}

\section{MAGNETIC PROPERTIES}  
\blue{We use $C_p(T)$,  $\chi(T)$, and $M(H)$ measurements to characterize the spin-anisotropy of $R$V$_6$Sn$_6$ and identify possible magnetic orderings. 
For each compound,  $\chi(T)$ and $M(H)$ measurements were carried out under external magnetic field applied both parallel and perpendicular to $c$-axis [Fig. \ref{Fig:Magpro}]. Modified C-W fit to $1/\chi(T)$ were performed at high (100-300 K) and low temperatures (6-10 K) to extract the effective moment $\mu_\textrm{eff}$, and C-W temperature ($\theta_{CW}$), respectively. For all compounds and field applied in both directions, the numbers for $\mu_\textrm{eff}$ are generally in agreement with the theoretical free-ion moment ($\mu_\textrm{theory}$) expected for R$^{3+}$ ions [Table \ref{table:all}], consistent with the localized moment picture of R$^{3+}$  magnetism.}

\subsection{TbV$_6$Sn$_6$} 
The $C_\textrm{mag}$ of TbV$_6$Sn$_6$ clearly shows a sharp peak at 4.4 K which can be seen clearly in $C_\textrm{mag}$ [blue symbols in Fig. \ref{Fig:HC}(b)]. A broad anomaly maximized around 40 K shows up in $C_\textrm{mag}$ at high temperature, which we attribute to the Schottky anomaly of Tb$^{3+}$ elevated crystal electric field. The integrated magnetic entropy $S_\textrm{mag}$ reaches $R\ln{2}$ at 10 K, and is relatively flat until 20 K, indicating that the lowest crystal field level is a non-Kramers doublet that is well separated from the higher levels. The fully recovered $S_\textrm{mag}$ from 2 K to 100 K reaches  21.77 J/mol-K, which is close to the full single-ion magnetic entropy of $R\ln{13}$  expected for Tb$^{3+}$ (total angular momentum $J = 6$).

The $\chi(T)$ of TbV$_6$Sn$_6$  exhibits paramagnetic behavior at high temperatures, \blue{while at low temperature, a broad anomaly in $\chi_{ab}$ is observed around 60 K which might be related to the broad $C_\textrm{mag}$ anomaly around 50 K. A zero-field cooling and field-cooling divergence is observed below 4.4 K  [Fig. \ref{Fig:Magpro}(a) inset], consistent with the sharp peak seen in $C_{mag}$.}  It is noteworthy that the absolute value of $\chi_{c}$ is two orders of magnitude larger than that of $\chi_{ab}$, indicating that the Tb magnetic moments tend to align along the crystallographic $c$-axis. \blue{This is further confirmed by the $M(H)$ at 0.4 K [Fig. \ref{Fig:Magpro}(f)], where the magnetization along $c$-axis rapidly increases and  saturates above 0.5 T to a moment of 9.72 $\mu_B$/Tb while the number is merely 1.25 $\mu_B$/Tb when the field is applied within $ab$-plane. 
With strong easy-axis anisotropy, a modified C-W fit of $\chi_{c}$ at low temperature yields $\theta_{CW} = 2.06$ K, suggesting an overall weak ferromagnetic interaction. }


\subsection{DyV$_6$Sn$_6$ and HoV$_6$Sn$_6$}
\blue{Similar to that of TbV$_6$Sn$_6$, we can identify clear signature of magnetic ordering in DyV$_6$Sn$_6$ and HoV$_6$Sn$_6$ from $C_\textrm{mag}$, at 2.4 K and 2.5 K, respectively.   Interestingly, for DyV$_6$Sn$_6$, an additional transition at 3 K is present in $C_\textrm{mag}$ whose origin is subject to further investigations.}  Broad anomalies between 20 K and 50 K also show up in $C_\textrm{mag}$ at high temperature due to the Schottky anomaly of crystal field effects. For both samples, the integrated magnetic entropy $S_\textrm{mag}$ \blue{continuously increases above the magnetic ordering temperature, whose numbers quickly exceed $R\ln{2}$. This observation clearly demonstrates the existence of low-lying crystal field, meaning that the low temperature rare earth magnetism cannot be treated as effective spin-1/2.} The fully recovered $S_\textrm{mag}$ from 2 K to 100 K reaches 25.52 J/mol-K for DyV$_6$Sn$_6$ and 24.51 J/mol-K for HoV$_6$Sn$_6$, which are in reasonable agreement with the full single-ion magnetic entropy of $R\ln{15}$ and $R\ln{16}$  expected for Dy$^{3+}$ ($J = 15/2$) and  Ho$^{3+}$ ($J = 8$), respectively.     

\begin{figure}
	\linespread{1}
	\par
	\begin{center}
		\includegraphics[width= 3.5 in ]{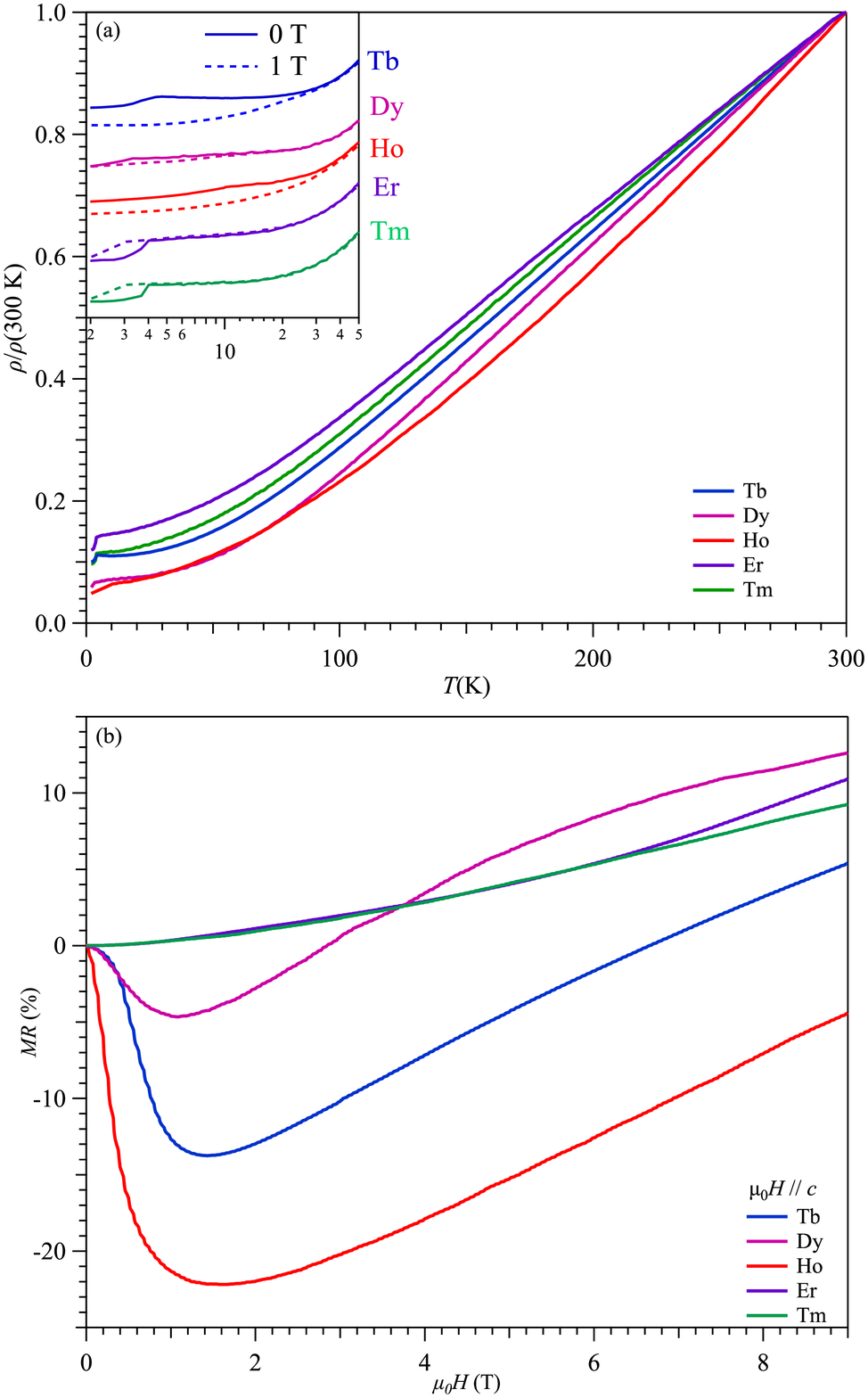}
	\end{center}
	\par
	\caption{\label{Fig:RTRH}  (a) The temperature dependence of normalized resistivity $\rho$/$\rho$(300 K) of $R$V$_6$Sn$_6$ ($R$ = Tb - Tm), insert: temperature dependence of resistivity measured without and with magnetic field of 1 T along $c$-axis. (b) The magnetic field dependence of magneto-resistivity of $R$V$_6$Sn$_6$ with external magnetic field applied along $c$-axis. }
\end{figure}

For DyV$_6$Sn$_6$, an abrupt anomaly in $\chi_{c}$ was observed at around 3 K while $\chi_{ab}$ \blue{becomes flat} below 2.4 K, consistent with the two sharp peaks seen in $C_{mag}$. For HoV$_6$Sn$_6$, $\chi_c$ becomes flat around 2.5 K while a broad peak is observed in $\chi_{ab}$ at this temperature, consistent with the sharp peak seen in $C_{mag}$. It is noteworthy that the absolute value of $\chi_{c}$ and $\chi_{ab}$ are in the same order of magnitude for DyV$_6$Sn$_6$ and HoV$_6$Sn$_6$, indicating that the Dy and Ho magnetic moments tend to exhibit Heisenberg-like behavior. This is further confirmed by the $M(H)$ at 0.4 K [Fig. \ref{Fig:Magpro}(f) and (h)]. For DyV$_6$Sn$_6$, the magnetization along $c$-axis rapidly increases above 0.5 T and keeps flat until 6 T \blue{where a jump to another plateau of 9.52 $\mu_B$/Dy appears [Fig. \ref{Fig:RTRH}(g) inset].} Accordingly, the $M_{ab}$ reaches 6.73 $\mu_B$/Dy at 7 T. For HoV$_6$Sn$_6$, the magnetization along $c$-axis rapidly saturates above 0.3 T to a moment of 8.81 $\mu_B$/Ho while moment keeps increase to 9.71 $\mu_B$/Ho until 7 T when the field is applied with $ab$-plane. This indicates the single-ion magnetism is still anisotropic and weak easy-axis.
The low temperature C-W fit of DyV$_6$Sn$_6$ gives $\theta_{CW} = 1.29 K$ and 1.65 K for $\chi_{ab}$ and $\chi_{c}$, respectively, suggesting an overall weak ferromagnetic interaction for both direction. Accordingly, the same C-W fit to HoV$_6$Sn$_6$ yields $\theta_{CW}^{ab} = -10.73$ K and $\theta_{CW}^{c} = -0.66$ K, indicating \blue{the dominating magnetic interaction is antiferromagnetic between Ho$^{3+}$ moments in the $ab$-plane.} 


\subsection{ErV$_6$Sn$_6$ and TmV$_6$Sn$_6$}
\blue{Different from the three compounds discussed above, we found no sign of magnetic ordering down to 1.8 K in ErV$_6$Sn$_6$ or TmV$_6$Sn$_6$ from $C_\textrm{mag}$. Instead,  $C_\textrm{mag}$ of TmV$_6$Sn$_6$ shows a broad peak below 10 K, likely due to development of short-ranged magnetic correlations, which is however absent in ErV$_6$Sn$_6$.} The integrated magnetic entropy $S_\textrm{mag}$ keeps increasing \blue{with neither anomaly nor plateaus observed until 60 K, suggesting the energy scale of the crystal fields are in the order of several meV}. For ErV$_6$Sn$_6$ and TmV$_6$Sn$_6$, the fully recovered $S_\textrm{mag}$ from 2 K to 100 K reaches  24.95 J/mol-K and 22.43 J/mol-K respectively, which is close to the full single-ion magnetic entropy of $R\ln{16}$ and $R\ln{13}$  expected for Er$^{3+}$($J = 15/2$) and Tm$^{3+}$ ($J = 6$).

For ErV$_6$Sn$_6$ and TmV$_6$Sn$_6$, there is no anomaly observed in $\chi_{ab}$, \blue{confirming the absence of magnetic ordering down to 0.4 K}. Broad anomalies in $\chi_{c}$ are observed around 20 K and 60 K which might be related to crystal field effects. No anomaly is observed in $\chi_{ab}$ and $\chi_{c}$ of ErV$_6$Sn$_6$ and TmV$_6$Sn$_6$ at low temperature, consistent with the hump in $C_{mag}$. It is noteworthy that the absolute value of $\chi_{ab}$ is two orders of magnitude larger than that of $\chi_{c}$ for these two compounds, indicating that the Er and Tm magnetic moments tend to align in $ab$-plane. This is further confirmed by the $M(H)$ at 0.4 K [Fig. \ref{Fig:Magpro}(f)]. For ErV$_6$Sn$_6$, the magnetization in $ab$-plane rapidly increases and saturate above 0.5 T to a moment of 7.41 $\mu_B$/Er while the moment keeps increasing to 3.39 $\mu_B$/Er at 7 T after  a transition appeared at around 0.4 T when the field is applied along $c$-axis. For TmV$_6$Sn$_6$, the magnetization in $ab$-plane rapidly reaches 6.29 $\mu_B$/Tm at 7 T, in sharp contrast to a number of 0.36 $\mu_B$/Er when the field is applied along $c$-axis \blue{, suggesting that the TmV$_6$Sn$_6$ can be properly described by an $XY$ effective spin model.} With easy-plane anisotropy for both compounds, a low temperature C-W fit of $\chi_{ab}$ gives $\theta_{CW}^{ab} = 0.07$ K  for ErV$_6$Sn$_6$, and $\theta_{CW}^{ab} = -3.18$ K for TmV$_6$Sn$_6$. \blue{This seems to suggest the average in-plane spin-spin interaction is antiferromagnetic for TmV$_6$Sn$_6$ while the magnitude is negligible in ErV$_6$Sn$_6$. }
 
\begin{figure*}[tbp!]
	\linespread{1}
	\par
	\begin{center}
		\includegraphics[width= 7 in]{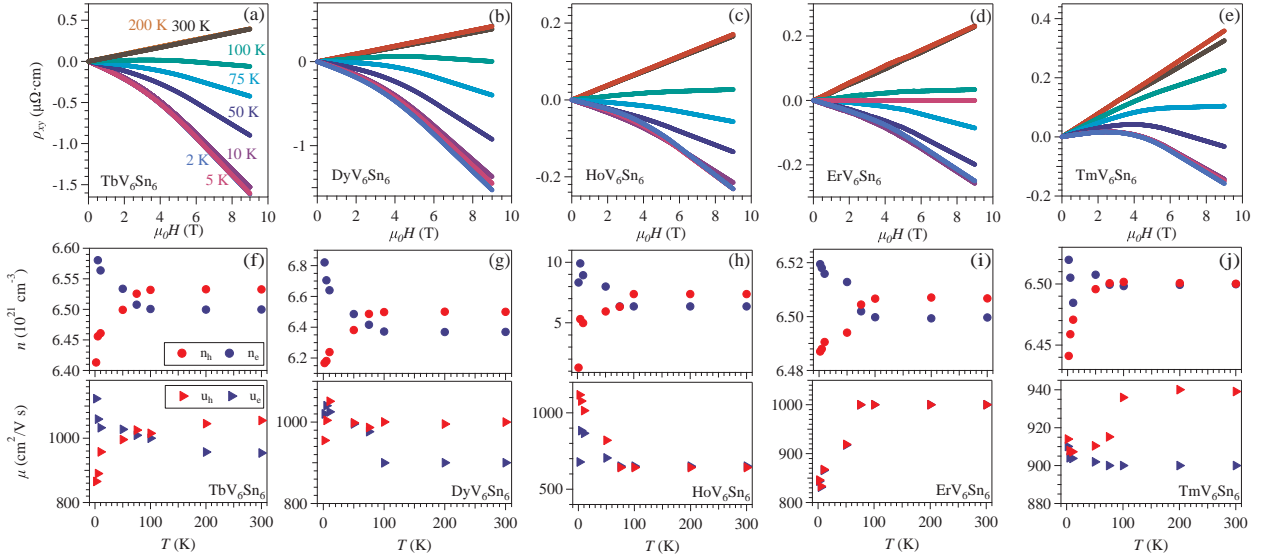}
	\end{center}
	\par
	\caption{\label{Fig:hall}  (a-e) Magnetic field dependence of longitudinal resistivity $\rho_{xy}$ measured at different temperatures. (f - j)Temperature dependence of carrier density,$n$ and mobility $\mu$ obtained from hall conductivity fit. }
\end{figure*}
\section{TRANSPORT PROPERTIES}
Fig.\ref{Fig:RTRH}(a) shows the temperature dependence of normalized resistivity $\rho(T)/\rho(300 K)$ at zero field for series $R$V$_6$Sn$_6$ ($R$ = Tb - Tm) compounds, and the typical metallic behavior can be seen for all samples. At low temperatures, an anomaly exhibits in the resistivity, which is associated with the formation of the magnetic order of $Tb^{3+}$, $Dy^{3+}$ and $Ho^{3+}$ ions. The inset of Fig.\ref{Fig:RTRH}(a) shows the enlarged view of resistivity in the low temperature range, the sharp drop in resistivity at the magnetic ordering temperature can be seen more clearly. For ErV$_6$Sn$_6$ and TmV$_6$Sn$_6$, anomaly was also observed at low temperature even without magnetic order. When we increase the magnetic field to 1 T, the anomaly was suppressed in resistivity. We thus attribute this consistent drop in $\rho(T)$ to the scattering of the conduction electrons by optical phonons if the magnetic s-f contribution is neglected \cite{RT2020}. The possibility of Sn impurity superconductivity cannot be ruled out either.
Here, we vertically shift all the resistivity curves for clarity. Moreover, the transport behavior has strong dependence to the $R$ ions, where we can see more clearly in the magneto-resistance (MR) at 2 K. As is shown in Fig.\ref{Fig:RTRH}(b), ErV$_6$Sn$_6$ and TmV$_6$Sn$_6$ exhibit a positive MR, while for $R$ = Tb, Dy and Ho, the negative MR is consistent with the formation of magnetic order of these $R$ ions in the magnetic measurements.
To further gain insights into the carrier information including carriers concentration and mobility of the series of $R$V$_6$Sn$_6$ samples, Hall resistivity measurements were performed. As displayed in Fig.\ref{Fig:hall}(a - e), the Hall resistivity $\rho_{xy}(H)$ were measured at various temperatures for the whole series of $R$V$_6$Sn$_6$ samples. The current was applied within the $ab$-plane and the magnetic field applied along the $c$-axis. The $\rho_{xy}(H)$ ($\rho_{xx}$) data were anti-symmetrized (symmetrized) with respect to the data collected between +9 and -9T. As can be seen, the $\rho_{xy}(H)$ for these compounds share similar behaviors with increasing temperature gradually. In specific, the $\rho_{xy}(H)$ curves are linear with hole-dominated single band feature at temperatures above 200 K and the non-linear behavior emerges at low temperatures, in agreement with the multi-band character. To obtain the carriers concentration and mobility in these compounds, two-band model was employed to fit the Hall conductivity $\sigma_{xy}(H)$ at different temperatures below 200 K. Here, the longitudinal resistivity $\rho_{xx}$ curves were also measured at the same time. The Hall conductivity $\sigma_{xy}(H)$ and two-carrier model were calculated based on the equation:

\begin{equation} \label{def of hallcon}
\mathcal \sigma_{xy} = - \frac{\rho_{xy}}{\rho_{xy}^2 + \rho_{xx}^2}
\end{equation}

\begin{equation}\label{FIT of hallcon}
\mathcal \sigma_{xy} = [n_h\mu_h^2\frac{1}{1 + (\mu_hB)^2} - n_e\mu_e^2\frac{1}{1 + (\mu_eB)^2}]eB
\end{equation}
where $n_e$ and $n_h$ are the carrier density of electrons and holes, while $\mu_e$ and $\mu_e$ are the mobility of electrons and holes.

Fig. \ref{Fig:hall}(f - j) show the fitting results of carriers density and mobility. Above 150 K, the density of hole for each sample is higher than that of electron. The hole mobility of TbV$_6$Sn$_6$, DyV$_6$Sn$_6$ and TmV$_6$Sn$_6$ is also dominant while electron and hole mobilities possess similar values for HoV$_6$Sn$_6$ and ErV$_6$Sn$_6$ above 150 K. It is also verified that $\rho_{xy}$ of all these samples exhibits hole-dominated behaviors.  At temperature below 100 K, a divergence for carrier density appears and seems to become stronger upon cooling down continuously. For DyV$_6$Sn$_6$, ErV$_6$Sn$_6$ and TmV$_6$Sn$_6$, the mobility of  electron and hole become comparable below 100 K. For TbV$_6$Sn$_6$, the electron carriers with larger mobility become dominant over the hole carriers while the carrier mobility of HoV$_6$Sn$_6$ has the opposite trend. In general, all sample exhibit hole-dominated behavior at high temperature while two band behavior prevails below 100 K, consistent with the  $\rho_{xy}$ data shown in Fig. \ref{Fig:hall}(a-e).  
~\\
~\\
{\centerline{\textbf{DISCUSSION AND SUMMARY}}}
~\\
~\\
In this study, we synthesized a series of HfFe$_6$Ge$_6$-type Kagome metals $R$V$_6$Sn$_6$ ($R$ = Tb - Tm) single-crystals and characterized their physical properties. Compared to the Kagome superconductors AV$_3$Sb$_5$ (A = K, Rb, Cs), no CDW or superconductivity were observed down to 0.4 K in $R$V$_6$Sn$_6$ ($R$ = Tb - Tm). 

In comparison with $R$Mn$_6$Sn$_6$, these V-based Kagome metals have similar evolution of spin anisotropy for the rare-earth elements as the their Mn-based counterparts \cite{ma2021PRLMn-based}. Specifically, TbV$_6$Sn$_6$ exhibits strong \blue{Ising} anisotropy, for which the magnetic moments tend to align to the $c$-axis while TmV$_6$Sn$_6$ and ErV$_6$Sn$_6$ possess \blue{easy-plane} anisotropy with moments lying in the $ab$-plane at low temperature. For $R$ = Ho, Dy,  more isotropic behaviors are observed. \blue{The dramatically different spin anisotropies originate from different crystal-field schemes of  R$^{3+}$ ions, whose effects further mediate the exchange couplings between 4f and 3d electrons, and thus becomes the key to engineer the magnetic structure and topological properties of $R$Mn$_6$Sn$_6$ \cite{ma2021PRLMn-based,yin2020NATURETb,gao2021AHE}. Moreover, the $XY$ anisotropy of Tm$^{3+}$ in TmMn$_6$Sn$_6$ is very surprising since all non-Kramers doublets should be  described by Ising moment of effective spin-1/2 \cite{Dun2021}, like that found in TbMn$_6$Sn$_6$. This means the magnetism of  TmMn$_6$Sn$_6$ must involve more than one crystal field levels and request a different theoretical treatment.  The absence of magnetic order in  TmV$_6$Sn$_6$ in contrast to a moderate spin-spin coupling ($\theta_{CW}^{ab}$ = -3.18 K) also indicates the existence of magnetic frustration, which is commonly found in insulating triangular lattice antiferromagnets. 
Consider the complexity of long-ranged Ruderman-Kittel-Kasuya-Yorsida, it calls for further investigations to understand the absence of magnetic order in TmV$_6$Sn$_6$.}

It has been reported that YV$_6$Sn$_6$ and GdV$_6$Sn$_6$ show qualitatively similar band structures in the paramagnetic state\cite{pokharel2021YGd}. As predicted by the density functional theory, the density of states in these materials are dominated by V $d$ states. Thus our new synthesized materials, $R$V$_6$Sn$_6$ ($R$ =Tb - Tm), are expected to possess similar electronic band structure. \blue{Indeed, our fits with two band model have revealed the multi-band nature of this family of material, hinting for non-trivial topological properties}.   

\begin{acknowledgements}
This work is supported by the National Key R\&D Program of China (2018YFA0305700, 2021YFA1400200), the National Natural Science Foundation of China (12025408, 11874400, 11834016, 11921004, 12174424), Beijing Natural Science Foundation (Z190008), the Strategic Priority Research Program and Key Research Program of Frontier Sciences of CAS (XDB25000000, XDB33000000 and QYZDB-SSW-SLH013), the CAS Interdisciplinary Innovation Team (JCTD-2019-01) and Lujiaxi international group funding of K. C. Wong Education Foundation (GJTD-2020-01)
\end{acknowledgements}

\bibliography{RV6Sn6}

\end{document}